\documentclass{ws-ijmpa}

\begin{document}

\markboth{J.~S.~Lange, Belle Collaboration}
{Charmed Hadrons from Fragmentation and $B$ Decays}

%\catchline{}{}{}{}{}

\title{Charmed Hadrons from Fragmentation and $B$ Decays}

\author{Jens S\"oren Lange, for the Belle Collaboration}

\address{
Johann Wolfgang Goethe-Universit\"at, Institut f\"ur Kernphysik,\\ 
Max-von-Laue-Stra\ss{}e 1, D-60438 Frankfurt am Main\\
\footnote{Present Address: Justus-Liebig-Universit\"at Gie\ss{}en, 
II.\ Physikalisches Institut, Heinrich-Buff-Ring 14, 35392 Gie\ss{}en,
Email Soeren.Lange@physik.uni-giessen.de}
}

\renewcommand\floatpagefraction{.99}
\renewcommand\topfraction{.99}
\renewcommand\bottomfraction{.99}
\renewcommand\textfraction{.01}   
\setcounter{totalnumber}{100}
\setcounter{topnumber}{100}
\setcounter{bottomnumber}{100}

\setlength{\floatsep}{1ex}
\setlength{\textfloatsep}{1ex}
\setlength{\intextsep}{1ex}
\setlength{\dbltextfloatsep}{1ex}
\setlength{\dblfloatsep}{1ex}
\setlength{\abovecaptionskip}{0cm}
\setlength{\belowcaptionskip}{0.0cm}

%%\author{for the Belle Collaboration}

\maketitle

%\begin{history}
%\received{Day Month Year}
%\revised{Day Month Year}
%\end{history}

\vspace*{-0.7cm}

\begin{abstract}
The fragmentation functions of 
$D^0$, $D^{\pm}$, $D_s^{\pm}$, $D^{*o}$, $D^{*\pm}$ and $\Lambda_c^{\pm}$
at $\sqrt{s}$$\simeq$10.6~GeV are measured with a data set
of 102.7~fb$^{-1}$.
Fragmentation model parametrizations 
(Peterson, Kartvelishvili, Collins-Spiller, Lund, and Bowler models)
are compared to the data.
The data at high $x$$\simeq$1 indicate a contribution 
of non-perturbative QCD processes.
\end{abstract}

\ccode{PACS numbers: 13.66.Bc, 13.87.Fh, 14.40.Lb}

%%%%%%%%%%%%%%%%%%%%
%%\vspace*{-0.3cm}
\section{Introduction}
%%%%%%%%%%%%%%%%%%%%

\noindent
The Belle experiment\cite{Belle} at the asymmetric $e^+$$e^-$ collider\cite{KEK-B}
KEK-B in Tsukuba, Japan, is operating with an $e^-$ beam energy
of 8.0~GeV and an $e^+$ beam energy of 3.5~GeV. 
The center-of-mass (cms) energy is adjusted to a 
$\sqrt{s}$=10.58~GeV, corresponding to the mass of the $\Upsilon(4S)$
resonance. At this cms energy, there is a resonant and
a non-resonant component: the resonant $\Upsilon(4S)$ is produced 
with a cross section of $\sigma$$\simeq$1.2~nb, the non-resonant 
con\-ti\-nu\-um with $\sigma$$\simeq$3.0~nb. 
The $\Upsilon(4S)$ decays into $B$$\overline{B}$ meson pairs
with a branching fraction higher than 99\%, and $\simeq$99\% 
of all $B$($\overline{B}$) mesons
decay into final states with hadrons containing charm quarks.
In the con\-ti\-nu\-um, the cross section of direct charm production 
is also high, i.e.\ $\sigma$$\simeq$1.3~nb. 
For about 10\% of the time, data are recorded at 
$\sqrt{s}$=10.52~GeV, containing only the non-resonant con\-ti\-nu\-um component.

\noindent
Fragmentation is a QCD process which involves 
not only single partons, but a cascade of many partons,
as it is shown schematically in Fig.~1.
A single bare charm quark starts radiating gluons.
The gluons themselves may 
a.) radiate additional gluons directly, or 
b.) convert into 
$u$$\overline{u}$, $d$$\overline{d}$ or $s$$\overline{s}$ 
quark-antiquark pairs, while those quarks then might radiate 
additional gluons. 
If the distances between the quarks become too large 
($r$$\geq$1~fm), hadronisation starts and final state
mesons and baryons are generated.

\begin{figure}[h]
\includegraphics[width=0.62\textwidth]{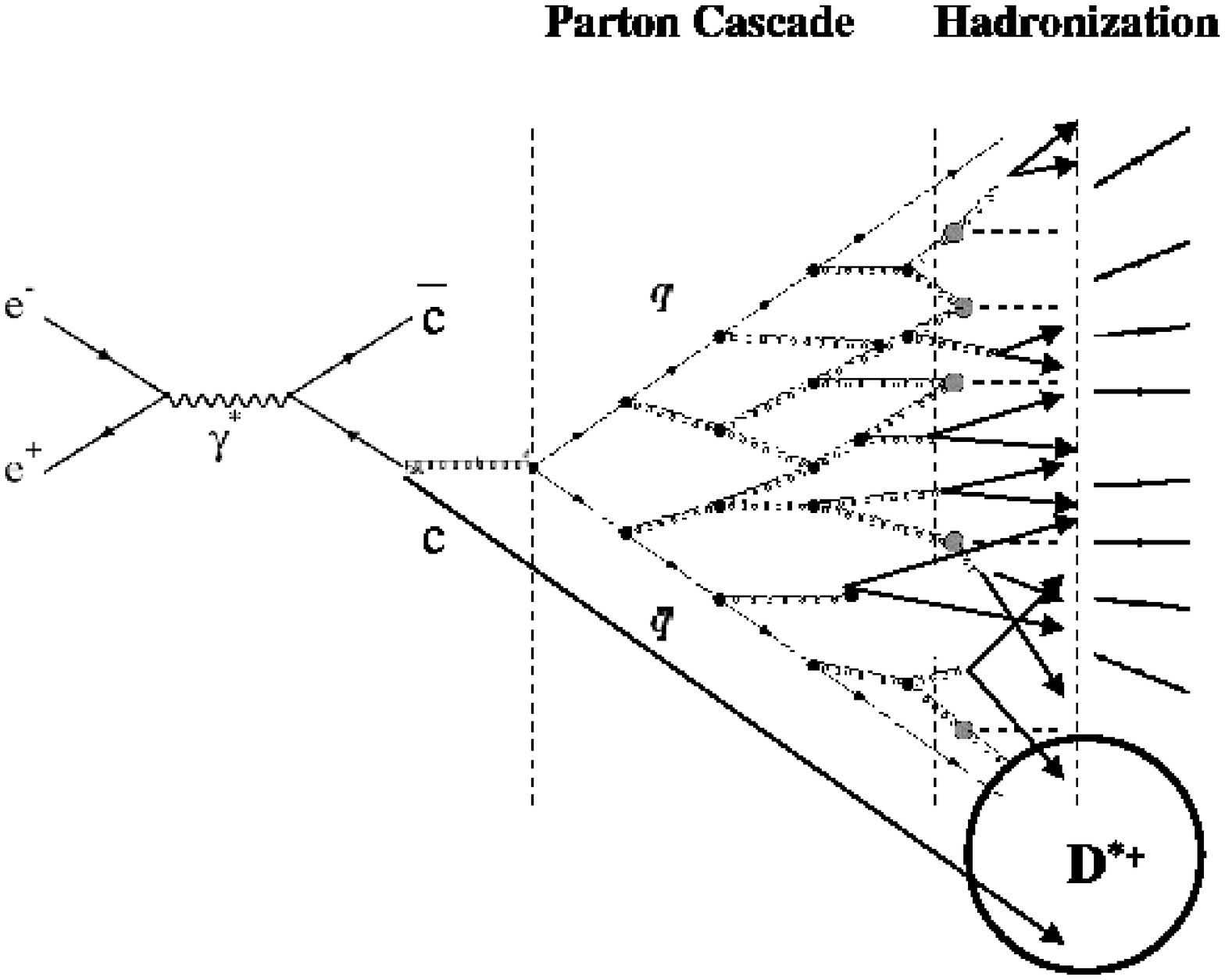}
\footnotesize
\vspace*{-2cm}
\begin{tabular}{@{}l@{}} 
Table 1. Decay modes used in the\\ 
reconstruction.\\
\toprule
$D^0$$\rightarrow$$K^-$$\pi^+$\\
$D^+$$\rightarrow$$K^-$$\pi^+$$\pi^+$\\
$D_s^+$$\rightarrow$$\Phi$$\pi^+$($\Phi$$\rightarrow$$K^+$$K^-$)\\
$\Lambda_c^+$$\rightarrow$$p$$K^-$$\pi^+$\\
$D^{*+}$$\rightarrow$$D^0$$\pi^+$($D^0$$\rightarrow$$K^-$$\pi^+$)\\
$D^{*+}$$\rightarrow$$D^+$$\pi^o$($D^+$$\rightarrow$$K^-$$\pi^+$$\pi^+$)\\
$D^{*o}$$\rightarrow$$D^0$$\pi^o$($D^0$$\rightarrow$$K^-$$\pi^+$)\\
\botrule
\quad \\
\quad \\
\quad \\
\quad \\
\quad \\
\quad \\
\quad \\
\quad \\
\quad \\
\quad \\
\quad \\
\quad \\
\quad \\
\quad \\
\quad \\
\quad \\
\quad \\
\quad \\
\quad \\
\quad \\
%\quad \\
%\quad \\
%\quad \\
\end{tabular} 
\vspace*{-3cm}
\caption{Schematical description of the process 
$e^+$$e^-$$\rightarrow$$c$$\overline{c}$ \hspace*{5.3cm}}
with subsequent fragmentation
of the charm quark into a $D^*$.
\end{figure}

\normalsize

%\begin{table}[ph]
%\tbl{Decay modes used in the reconstruction.}
%{\begin{tabular}{@{}l@{}} \toprule
%$D^0$$\rightarrow$$K^-$$\pi^+$\\
%$D^+$$\rightarrow$$K^-$$\pi^+$$\pi^+$\\
%$D_s^+$$\rightarrow$$\Phi$$\pi^+$($\Phi$$\rightarrow$$K^+$$K^-$)\\
%$\Lambda_c^+$$\rightarrow$$p$$K^-$$\pi^+$\\
%$D^{*+}$$\rightarrow$$D^0$$\pi^+$($D^0$$\rightarrow$$K^-$$\pi^+$)\\
%$D^{*+}$$\rightarrow$$D^+$$\pi^o$($D^+$$\rightarrow$$K^-$$\pi^+$$\pi^+$)\\
%$D^{*o}$$\rightarrow$$D^0$$\pi^o$($D^0$$\rightarrow$$K^-$$\pi^+$)\\
%\end{tabular} 
%\label{t_1}}
%\end{table}

\noindent
The study of fragmentation
has two aspects:
On the one hand, as a theoretical aspect, 
QCD predicts hadronic confinement,
but does not predict
how an initial state quark as a pointlike object
transforms into a final state hadron, 
i.e.\ an confined object with a formfactor.
In most fragmentation models, the splitting of one parton to many partons (the ``parton shower'') 
is treated in perturbative QCD only, and hadronisation models are only
phenomenological with minor predicting capabilities.
Heavy quarks are in particular of interest, because their production
is strongly suppressed in both the parton shower and the hadronisation.
Thus, a final state charmed hadron contains a charm quark, which was produced 
with a high probability in the primary interaction.

\noindent
On the other hand, as a practical aspect, 
it is important to improve Monte-Carlo (MC) event generators.
Several other analyses of direct charm production e.g.\ production of 
$D^{**}$, $D_{sJ}$ etc.\ require MC simulations
of charm fragmentation.
In this paper PYTHIA\cite{PYTHIA} 6.2 is used.
The event generators use parameters, which have to be constrained 
by comparison to experimental data.

\noindent
The relevant variable for the investigation of the process dynamics 
is the momentum fraction
$x$=$p_{hadron}$/$p_{quark}$.
As the quark momentum is a priori unknown and can only be approximated, 
in this paper $x_p$=$p$/$p_{max}$ is used, 
where $p_{max}$ is corrected for the mass of the final state hadron,
i.e.\ $p_{max}$=$\sqrt{s/4-m^2_{hadron}}$.
The fragmentation function $D(x_p)$ is then given as the normalized
yield as a function of $x_p$. Instead of the yield, 
related quantities such as the cross
section d$\sigma$/d$x_p$ or BR$\times$d$\sigma$/d$x_p$ can be used,
where BR denotes the branching ratio into the particular final state.
For the functional form of fragmentation functions, 
in PYTHIA the variable $z$ is used instead of the variable $x_p$,
which is defined as $z$=($E$+$p_z$)$_{hadron}$/($E$+$p_z$)$_{quark}$.

\noindent
As will be shown below, in particular the high $x$ region (i.e.\
$x_p$$\simeq$1 or $z$$\simeq$1) might be sensitive to effects of 
non-perturbative QCD. The $x$=1 limit is corresponding to the exclusive process 
e.g.\ $e^+$$e^-$$\rightarrow$$c$$\overline{c}$$\rightarrow$$D^{0}$$\overline{D}^{0}$,
in which on either side a bare charm quarks transforms into a hadron, which
contains the identical charm quark as a valence quark, but additionally one 
light valence quark, light sea quarks and sea gluons. The interesting aspect is that, due to 
$p_{hadron}$=$p_{quark}$, no momentum is available to create the
hadronic structure, but in fact the momentum balance is provided 
by the QCD vacuum.

\noindent
Fragmentation data of $e^+$$e^-$$\rightarrow$$c$$\overline{c}$ 
at $\sqrt{s}$$\simeq$10.6~GeV have been measured previously, 
but with lower statistical significance.
More than 10 years ago, the ARGUS collaboration and the CLEO collaboration
published\cite{argus_1}$^,$\cite{cleo_1} data based
upon integrated luminosities of 31.4~$pb^{-1}$ and 35.8~$pb^{-1}$, respectively.
Nowadays, equivalent data are recorded by BELLE in $\simeq$50~minutes.
More recently, the CLEO collaboration
published\cite{cleo_2} $D$ and $D^*$ fragmentation 
data based upon 8.9~$fb^{-1}$.
$D_s$ and $D_s^*$ fragmentation data were published\cite{cleo_3}$^,$\cite{babar}
by CLEO and by BABAR based upon 4.7~$fb^{-1}$ and 23.4~$fb^{-1}$, respectively.

%%%%%%%%%%%%%%%%%%
\vspace*{-0.3cm}
\section{Analysis}
%%%%%%%%%%%%%%%%%%

\noindent
A data set of 87.7~$fb^{-1}$ on-resonance 
and 15.0~$fb^{-1}$ off-resonance is used,
and compared to Monte Carlo (MC) simulations 
corresponding to 217.0~$fb^{-1}$.
The MC simulations are based upon 
the event generator QQ98\cite{QQ98} for $e^+$$e^-$$\rightarrow$$c$$\overline{c}$, 
the Peterson\cite{Peterson} fragmentation model,
PYTHIA\cite{PYTHIA} 6.2 for the parton shower evolution
and GEANT\cite{GEANT} 3.21 for the detector description.
For specific simulations, also different fragmentation models 
are used (see below). The decay modes used in the reconstruction 
are listed in Tab.~1.
Charge conjugated modes are included.
The efficiencies for particle identification of $\pi^{\pm}$ and $K^{\pm}$ are
above 96\% for all decay modes, the efficiency for protons
is 81\%.
The particle misidentification probabilities are between 12\% and 26\%.
This paper represents a short version of the complete analysis
which is published elsewhere\cite{prd}.
\noindent
With this unprecedented statistics it became possible 
to reduce the bin width in $x_p$ by a factor of $\simeq$3,
and thus investigate the shape of the fragmentation function
with high precision, in particular in the high $x_p$$\simeq$1 region.
In addition, fragmentation into the baryon $\Lambda_c$
is measured, for which a high statistics data set has not been 
available previously.

\noindent
Fig.~2 shows the mass distribution\footnote{The mass distributions in Fig.~2 
are shown as mass difference distributions of the measured mass and the mass
as given by the PDG\cite{PDG}.} for the $D^{0}$ and the $\Lambda_c^+$ 
(as an example for the meson and baryon reconstruction).
The signals are shown for a momentum fraction 0.68$<$$x_p$$<$0.70,
which is close to the $x_p$ of peak position in the fragmentation function.
The cross sections are calculated using 
the fitted signal yields.

\noindent
As mentioned above, at $\sqrt{s}$=10.6~GeV there is a resonant and a
non-resonant contribution. Thus, charmed hadrons can be produced by
$B$ decays (whereas the $B$ mesons are produced by decay of the 
$\Upsilon(4S)$ resonance) or by fragmentation. 
In a first step, these two contributions have to be identified.
Fig.~3 shows the cross section d$\sigma$/d$x_p$ as a 
function of $x_p$. The down-left hatched histogram shows on-resonance
data, the down-right hatched histogram shows off-resonance data.
After normalisation using the integrated luminosities of the respective samples, the na\"{\i}ve
$1/s$ dependence on the total hadronic cross section has been taken
into account by a weighting factor $(10.58~GeV/10.52~GeV)^2$.
The contribution of $B$ decays is dominant at lower $x_p$$<$0.5, as the $B$ decay 
kinematics limits the maximum possible momentum of the charmed hadron 
to about $p_{hadron}$$\simeq$2.5~GeV/c, corresponding to $x_p$$\simeq$0.5. 
The distributions in Fig.~3 are not yet efficiency corrected
and only statistical errors are shown.

%%%%%%%%%%%%%%%%%%%%%%%%%%%%%%%%%
\vspace*{-0.3cm}
\section{Fragmentation Functions}
%%%%%%%%%%%%%%%%%%%%%%%%%%%%%%%%%

\noindent
Fig.~4 shows the cross section d$\sigma$/d$x_p$ as a 
function of $x_p$.
For $x_p$$\leq$0.5 only off-resonance data are used.
For $x_p$$>$0.5 the weighted average of off-resonance and on-resonance data 
is used.
Fig.~4 is efficiency corrected, and represents the measured fragmentation 
functions $D(x)$, which can be compared to models.
The inner error bars show the statistical, 
the outer error bars the total uncertainties.
Fig.~4 is not yet corrected for feed-down from $D^*$ to $D^0$.
Details of the feed-down contribution are described elsewhere\cite{prd}.
The estimated systematic error from $D^{**}$ feeddown
(e.g.~$D_0^*$(2308) and $D'_1$(2427))
is included and is $\leq$13\%.
Tab.~\ref{t_2} shows the functional form of the different fragmentation
models which are considered. These are 
the Peterson\cite{Peterson},
the Kartvelishvili\cite{Kartvelishvili} 
the Collins-Spiller\cite{Collins-Spiller}
the Lund\cite{Lund}
and the Bowler\cite{Bowler} models.
The (1-$x$) term in all functional forms originates 
from the kinematics of radiation of one gluon.
For every model a $\chi^2$ is calculated from the bin-by-bin differences
between the model and the data, while the model parameters
are optimized in a fit procedure in order to find the optimum values.
Fig.~5 shows the measured $D^{*+}$ fragmentation function in comparison
to the models, and 
Tab.~\ref{t_2} lists the minimum $\chi^2$/{\it d.o.f.}.
The number of the degrees of freedom ({\it d.o.f.}) is given by the number
of $x_p$ bins minus the free parameters of the fit
(one or two, depending upon the model).
For the model comparison, the $D^*$ is chosen instead of the $D^0$ 
in order to avoid the feed-down correction.
The $\chi^2$ values for all other hadrons can be found elsewhere\cite{prd}.
The $\chi^2$ is not based upon the pure functional form, but instead
the MC generated fragmentation function according to the functional form.
Thus, the detector resolution, as modeled by GEANT, as well as the parton 
shower, as modeled by PYTHIA, is taken into account.
The models by Collins and Spiller and by Kartvelishvili are not included 
in the PYTHIA generator. Therefore, a reweighting procedure
has been applied. For each event, the $z$ and the $p_{\perp}$ values are
stored, and the fragmentation function ansatz is recalculated.
Tab.~\ref{t_3} shows the parameters at the minimum $\chi^2$ of the fit
for all different hadrons.
The fitted parameters indicate that the PYTHIA default parameters are not well 
suited for charm fragmentation. As an example, the PYTHIA default for the exponent
$a$ of the (1-$x$)$^a$ term in the Lund fragmentation model is $a$=0.30, but our 
fit gives as result $a$=0.68 at the minimum $\chi^2$. However, this is a global PYTHIA 
parameter, i.e.\ the parameter is valid for fragmentation of all quark flavors, 
and for e.g.\ the light quark sector the parameter might be different from our result.
As a result of the comparison, the Bowler and the Lund models appear favored.
This is consistent with an observation which was also reported\cite{SLD} 
for $B$ meson fragmentation. 
The reason might by given by the $exp$($-b$$m_{\perp}^2$/$z$) term
with the explicit use of the transverse mass $m_{\perp}$
which appears in the functional form of these two models, 
but not in any of the other models.
%\quad\\
%\quad\\
%\quad\\

%%%%%%%%%%%%%%%%%%%%%%%%%
\vspace*{-0.3cm}
\section{High $x$ region}
%%%%%%%%%%%%%%%%%%%%%%%%%

\noindent
The measured data were used in a theoretical study\cite{nonperb} 
in order to investigate possible non-perturbative QCD effects. 
For this purpose, the fragmentation function is split into a perturbative 
and a non-perturbative part, i.e.\ $D(x)$=$D_P(x)$+$D_{NP}(x)$, with 

\vspace*{-0.3cm}
\begin{equation}
D_{NP} = f_{NP} \cdot \frac{1}{1+c} [ \delta(1-x) 
+ c \frac{1}{N_2} (1-x)^a x^b ]
\end{equation}

\noindent
using the normalisation $N$=$\int_0^1$$(1-x)^a$$x^b$.
The factor $f_{NP}$ indicates the relative fraction of the non-perturbative
contribution and is determined by a fit.
While for the cases of the $D^*$ mesons the non-perturbative contribution
is $f_{NP}$$\simeq$21-25\%, for the $D^0$ meson 
it turns out to be quite high with 
$f_{NP}$$\simeq$60\%. The reason is, that a large fraction of the $D^0$ mesons 
arises from $D^*$ decay. The decay is very close to threshold, and thus
the hadron momenta are almost equal, i.e.\ $p$($D^*$)$\simeq$$p$($D^0$).
However, in the decay, a hadronic formfactor transition from a spin 1 hadron 
(i.e.\ the $D^*$) to a spin 0 hadron (i.e.\ the $D^0$) occurs. 
The partons involved into this transition must be soft, 
and thus non-perturbative QCD power corrections are expected to occur.

%%%%%%%%%%%%%%%%%
\vspace*{-0.3cm}
\section{Summary}
%%%%%%%%%%%%%%%%%

Charm fragmentation functions were measured at $\sqrt{s}$$\simeq$10.6~GeV 
with high precision. The data were compared to five different fragmentation 
models, and the model parameters for the best fit to the data were given.
The parameters can be used as input for PYTHIA. 
Among the five models, the Bowler and the Lund model are favoured.
The data are posted\cite{durham} to the DURHAM reaction database.

%%%%%%%%%%%%%%%%%%%%%%%%%%
%\section*{Acknowledgments}
%%%%%%%%%%%%%%%%%%%%%%%%%%

%We thank the KEKB group for excellent operation of the
%accelerator, the KEK cryogenics group for efficient solenoid
%operations, and the KEK computer group and
%the NII for valuable computing and Super-SINET network
%support.  We acknowledge support from MEXT and JSPS (Japan);
%ARC and DEST (Australia); NSFC and KIP of CAS (China); 
%DST (India); MOEHRD, KOSEF and KRF (Korea); 
%KBN (Poland); MIST (Russia); ARRS (Slovenia); SNSF (Switzerland); 
%NSC and MOE (Taiwan); and DOE (USA).

%\section{References}

\vspace*{-0.3cm}

%\begin{thebibliography}{000} %for 3 digits
%\begin{thebibliography}{00}  %for 2 digits

\clearpage

\begin{figure}[h]
\centerline{
\includegraphics[width=0.38\textwidth]{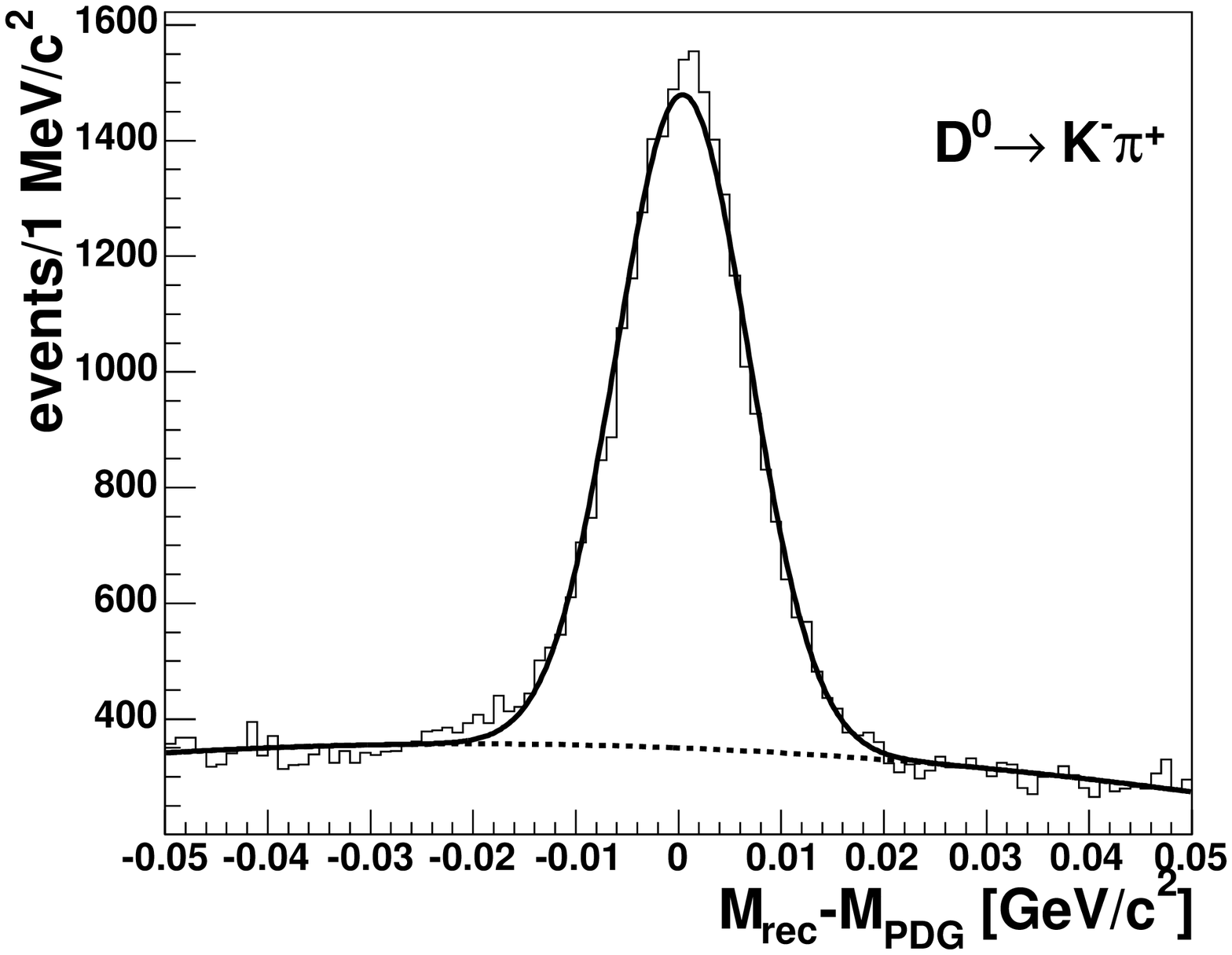}
\includegraphics[width=0.38\textwidth]{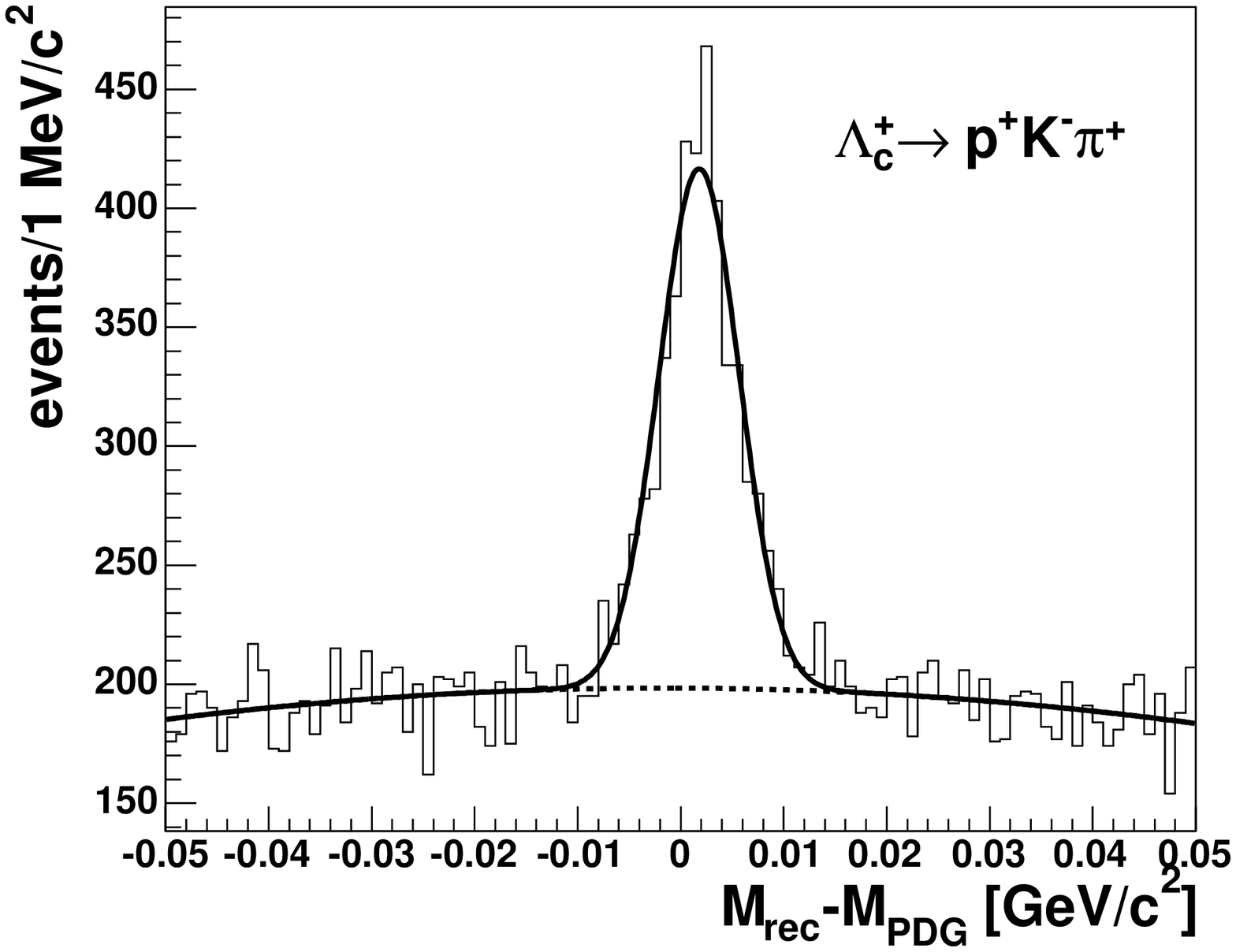}
}
\vspace*{-0.1cm}
\caption{
Mass distributions for the $D^{0}$ (left) and the $\Lambda_c^+$ (right)
for a momentum fraction 0.68$<$$x_p$$<$0.70.
The histograms show the data, the dotted lines describe only 
the background, the full lines describe the signal and the background.}
\end{figure}

\begin{figure}[h]
\centerline{
\includegraphics[width=0.38\textwidth]{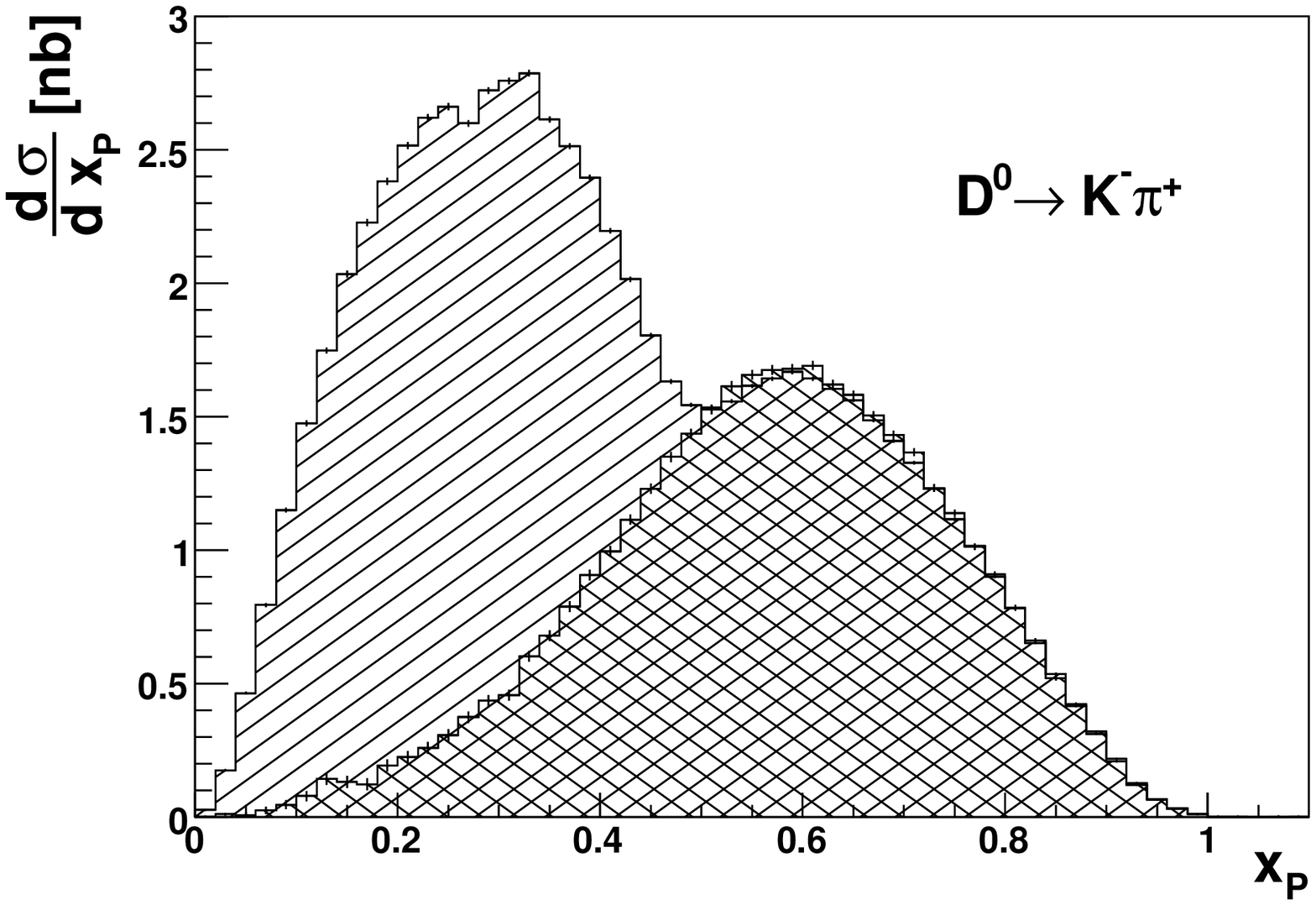}
\includegraphics[width=0.38\textwidth]{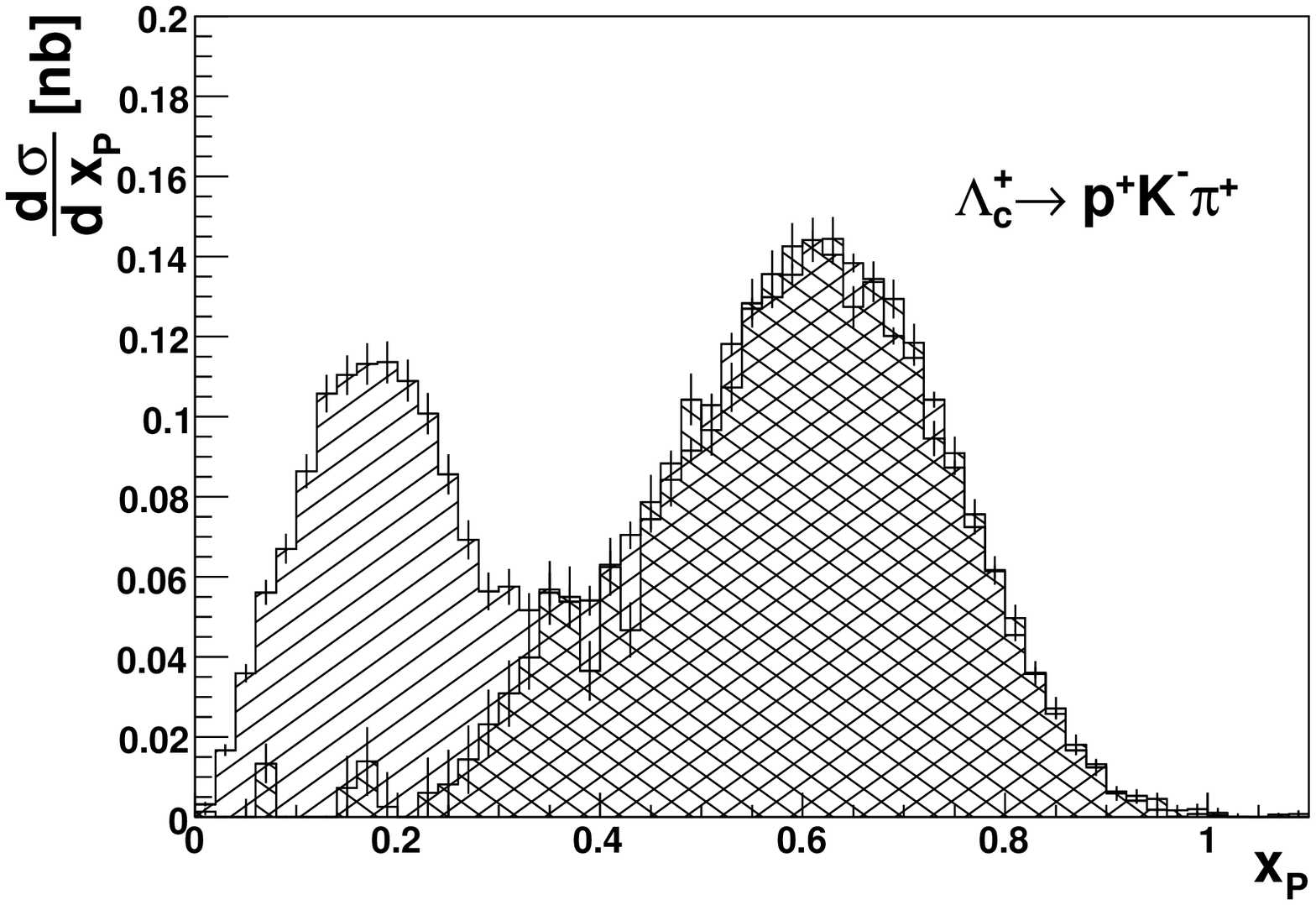}
}
\vspace*{-0.3cm}
\caption{The cross section d$\sigma$/d$x_p$ as a function of $x_p$
for $D^{0}$ (left) and $\Lambda_c^+$ (right). The down-left hatched 
histogram shows on-resonance data, the down-right hatched histogram 
off-resonance data.}
\end{figure}

\begin{figure}[h]
\centerline{
\includegraphics[width=0.38\textwidth]{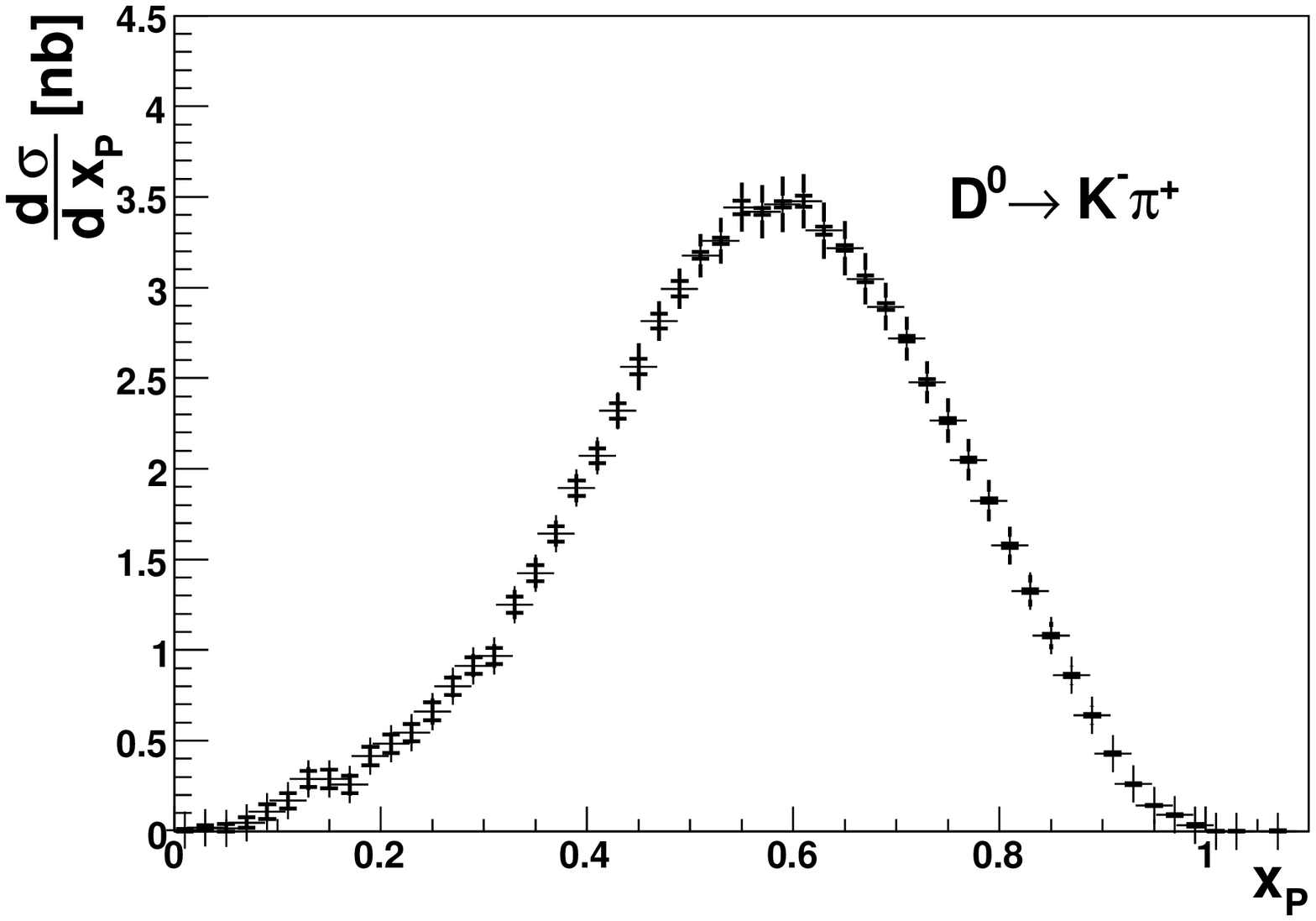}
\includegraphics[width=0.38\textwidth]{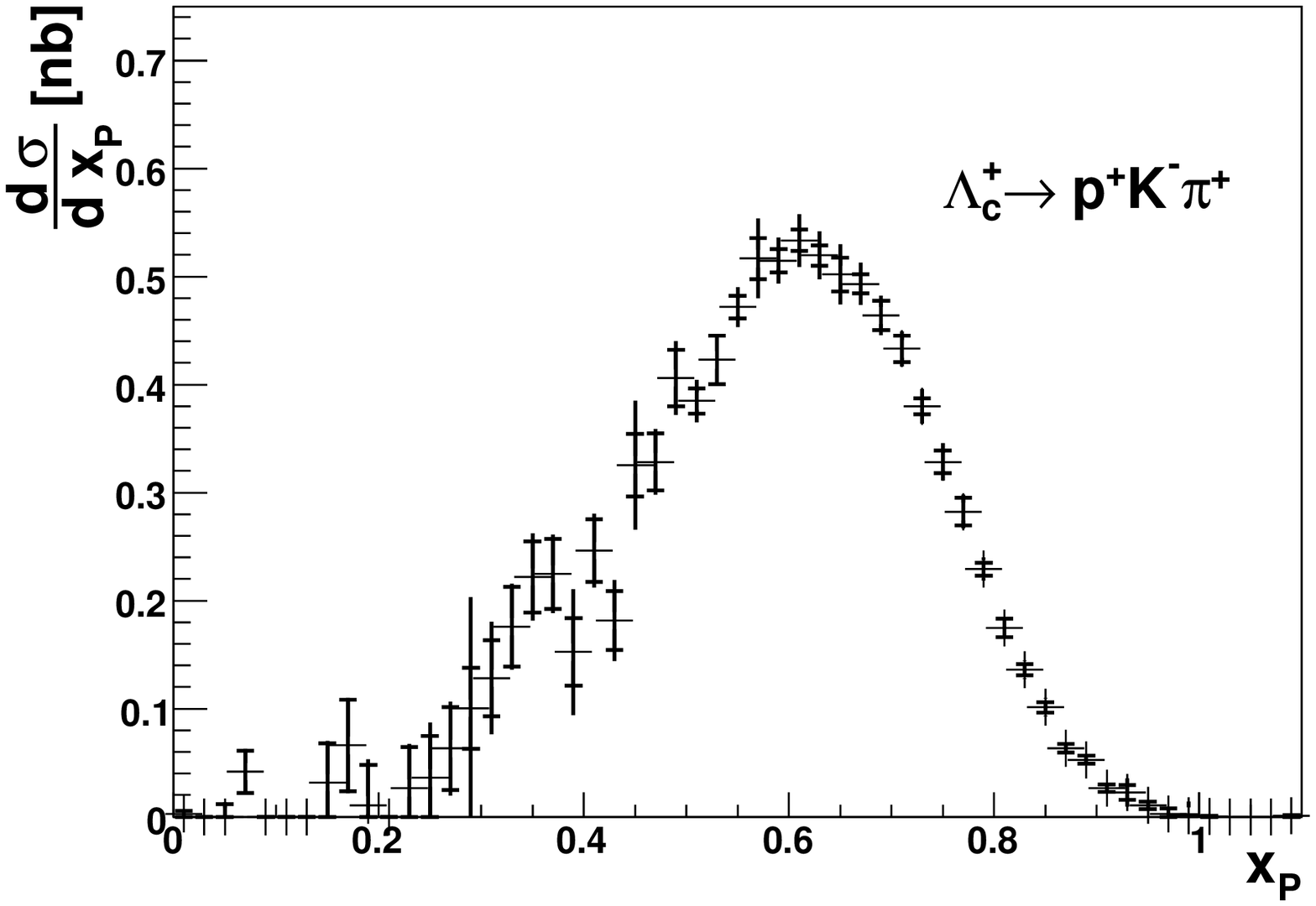}
}
\vspace*{-0.3cm}
\caption{Efficiency corrected cross section 
d$\sigma$/d$x_p$ as a function of $x_p$.
For $x_p$$\leq$0.5 only off-resonance data are used.
For $x_p$$>$0.5 the weighted average of off-resonance and on-resonance data 
is used.}
\end{figure}

\begin{table}[hhh]
\vspace*{0.2cm}
\tbl{Functional form of the fragmentation functions 
used in this analysis. The normalisation $N$ is different 
for all functions. In the last column, the $\chi^2$/d.o.f.
from the comparison to the data for the case of the $D^{*+}$ 
(see Fig.~5) is shown as an example.}
{\begin{tabular}{@{}lcr@{}} \toprule
Model & Functional Form & $\chi^2$/$d.o.f.$ for $D^{*+}$ \\ \hline
Bowler &
$N {1 \over z^{1+bm^2}} (1-z)^a \exp \left(-{b m_\perp^2 \over z}\right)$
& 541.8 / 55 \\
Lund &
$N {1 \over z} (1-z)^a \exp \left(-{b m_\perp^2 \over z}\right)$
& 965.6 / 55 \\
Kartvelishvili &
$Nz^{\alpha_{c}}(1-z)$  & 1271.1 / 54 \\                            
Collins-Spiller &
$N\left({1-z \over z} + {(2-z)\varepsilon_{c}' \over 1-z}\right)
(1+z^2) \left(1-{1 \over z}-{\varepsilon_{c}' \over
    1-z}\right)^{-2}$ & 1540.7 / 54 \\
Peterson &
$N {1 \over z}\left(1-{1\over z}-{\varepsilon_{c} \over
    1-z}\right)^{-2}$  & 3003.0 / 54 \\
\botrule
\end{tabular} 
\label{t_2}}
\end{table}

\clearpage

\begin{figure}[h]
\centerline{
\includegraphics[width=0.50\textwidth]{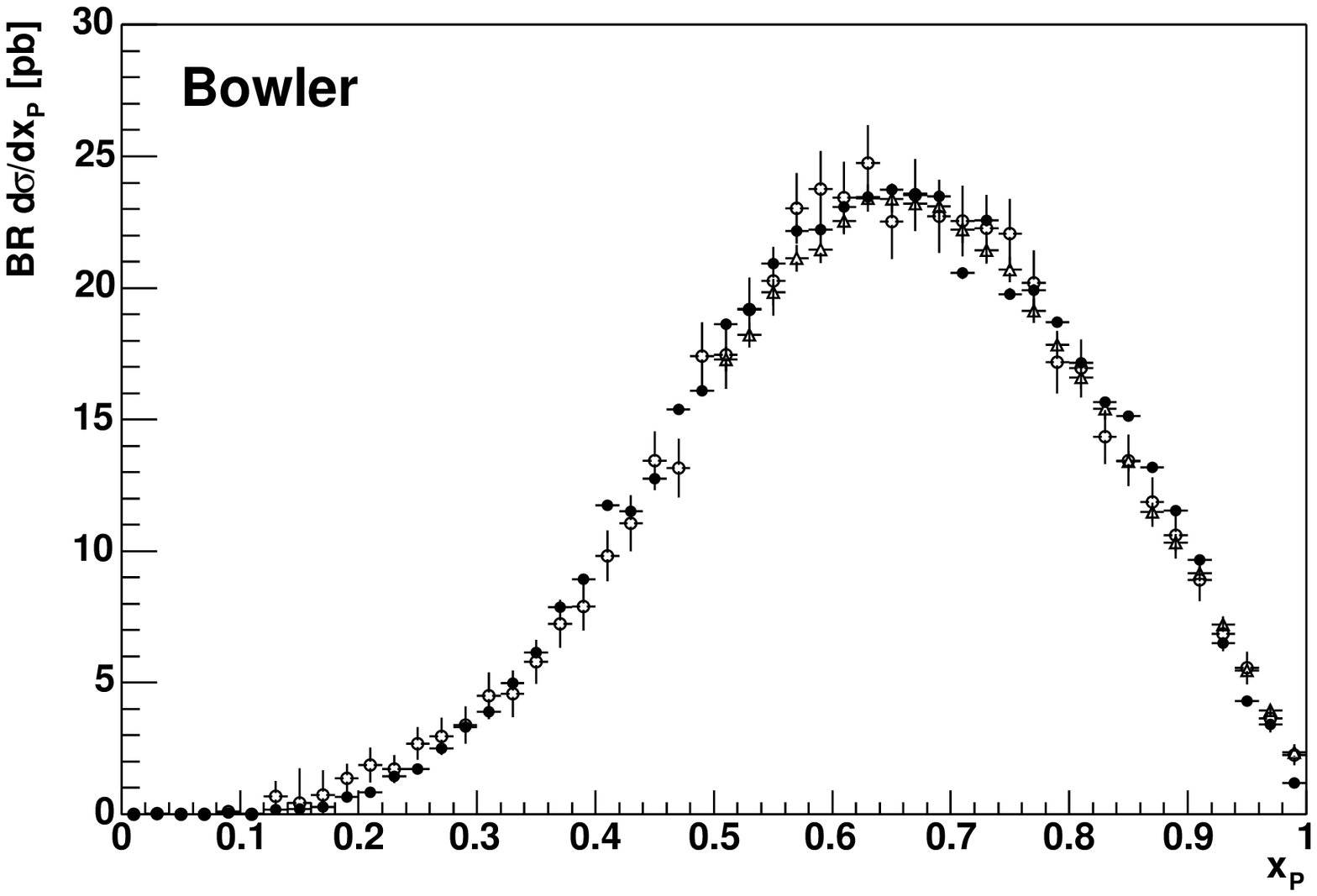}
\includegraphics[width=0.50\textwidth]{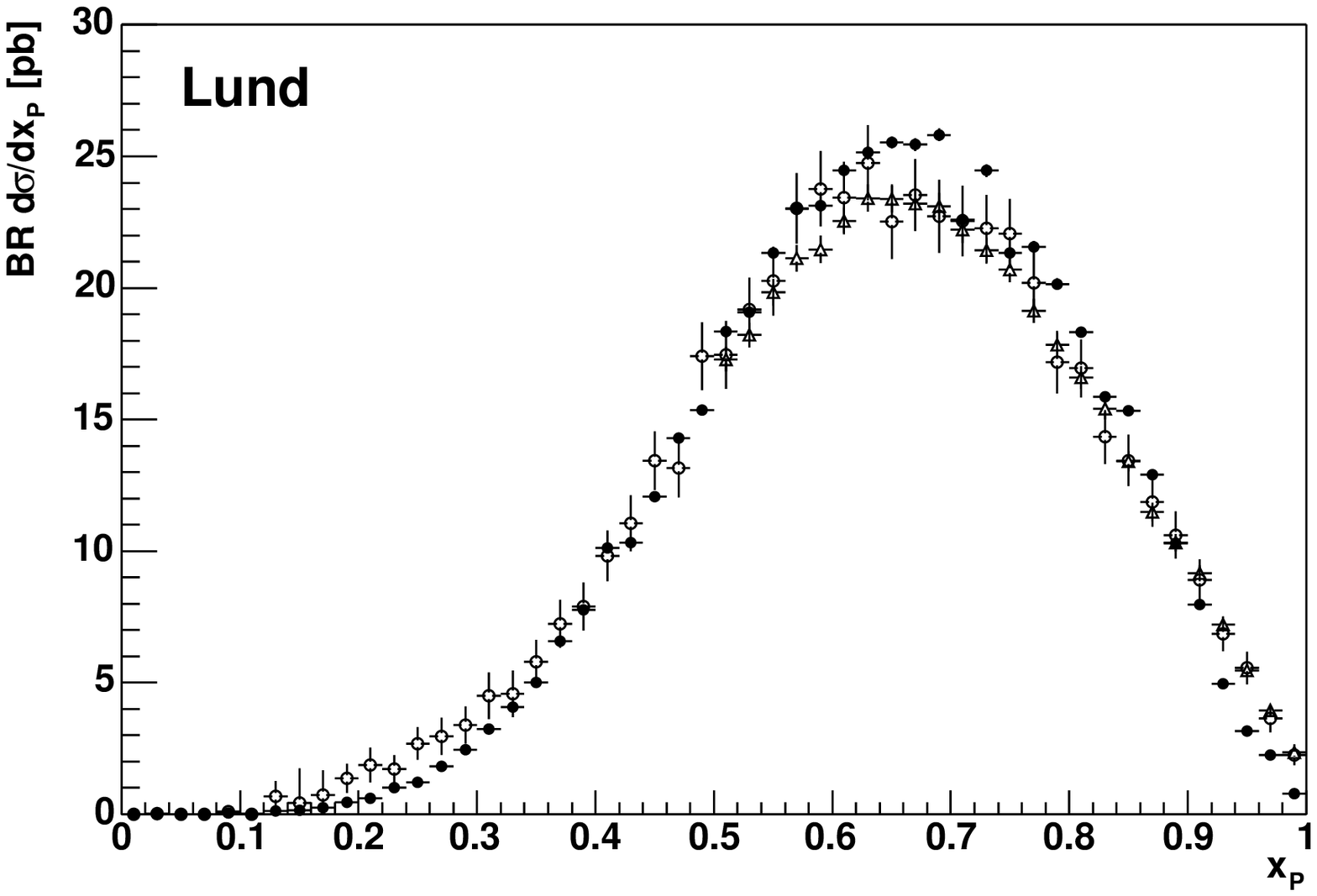}
}
\centerline{
\includegraphics[width=0.50\textwidth]{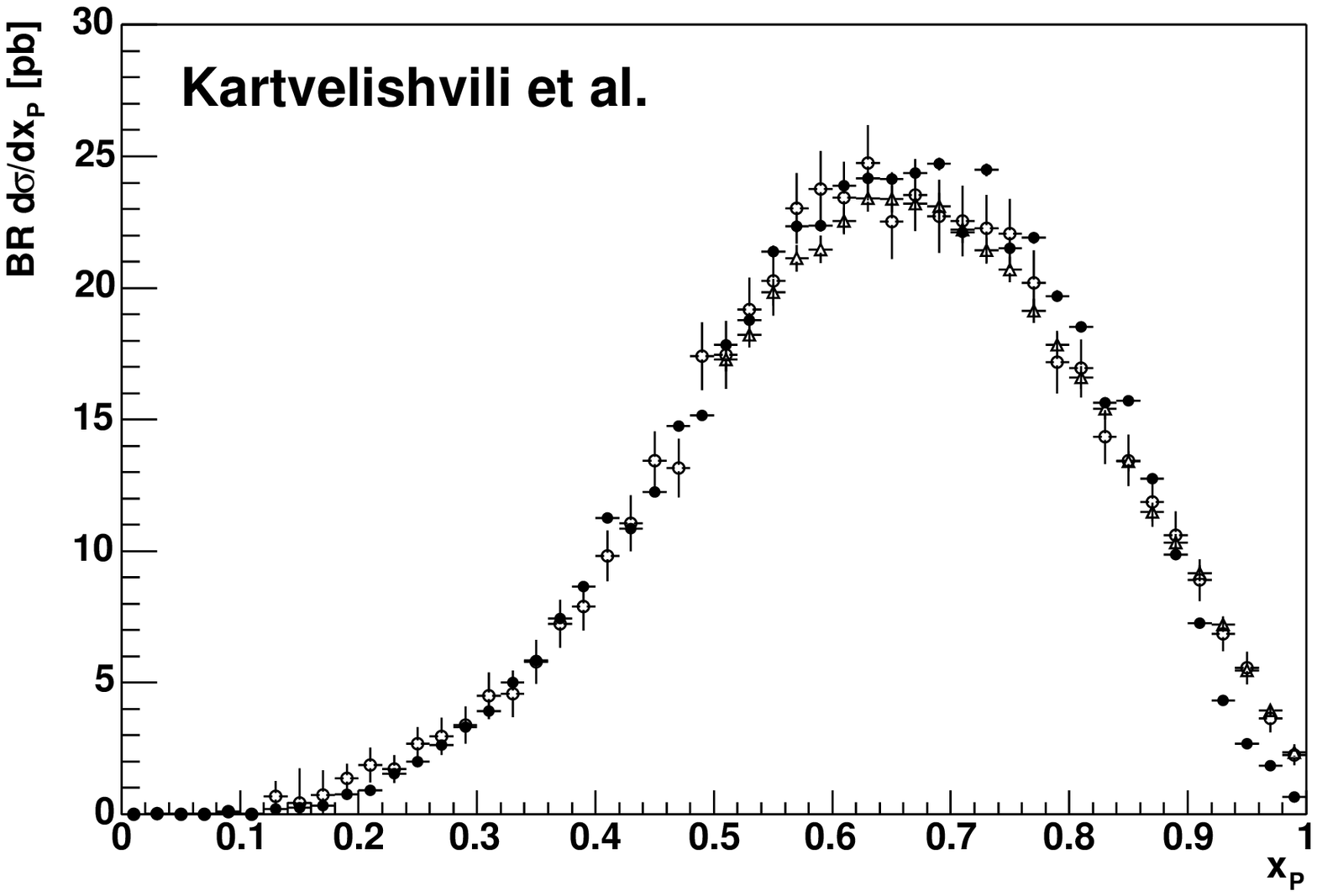}
\includegraphics[width=0.50\textwidth]{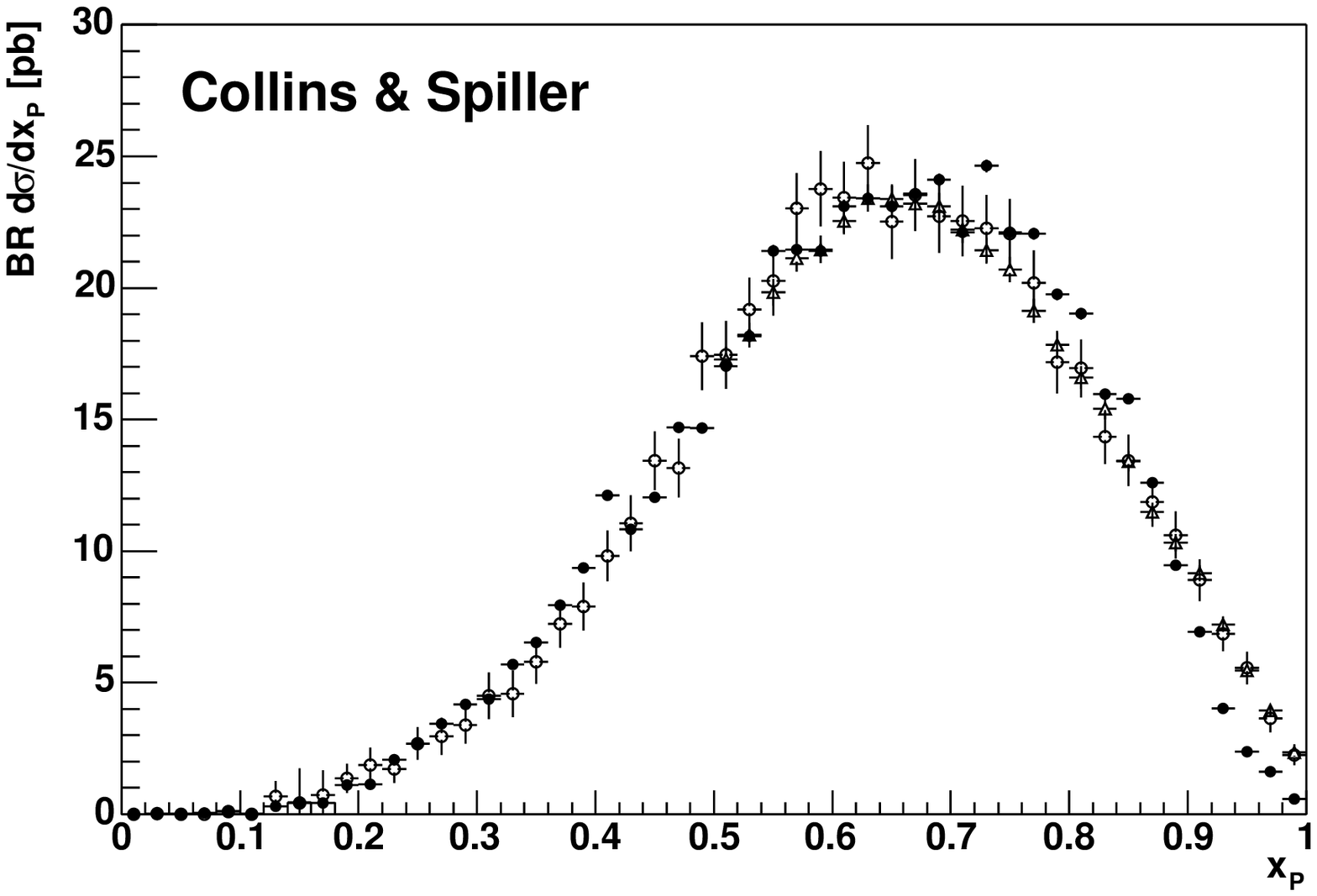}
}
\centerline{
\includegraphics[width=0.50\textwidth]{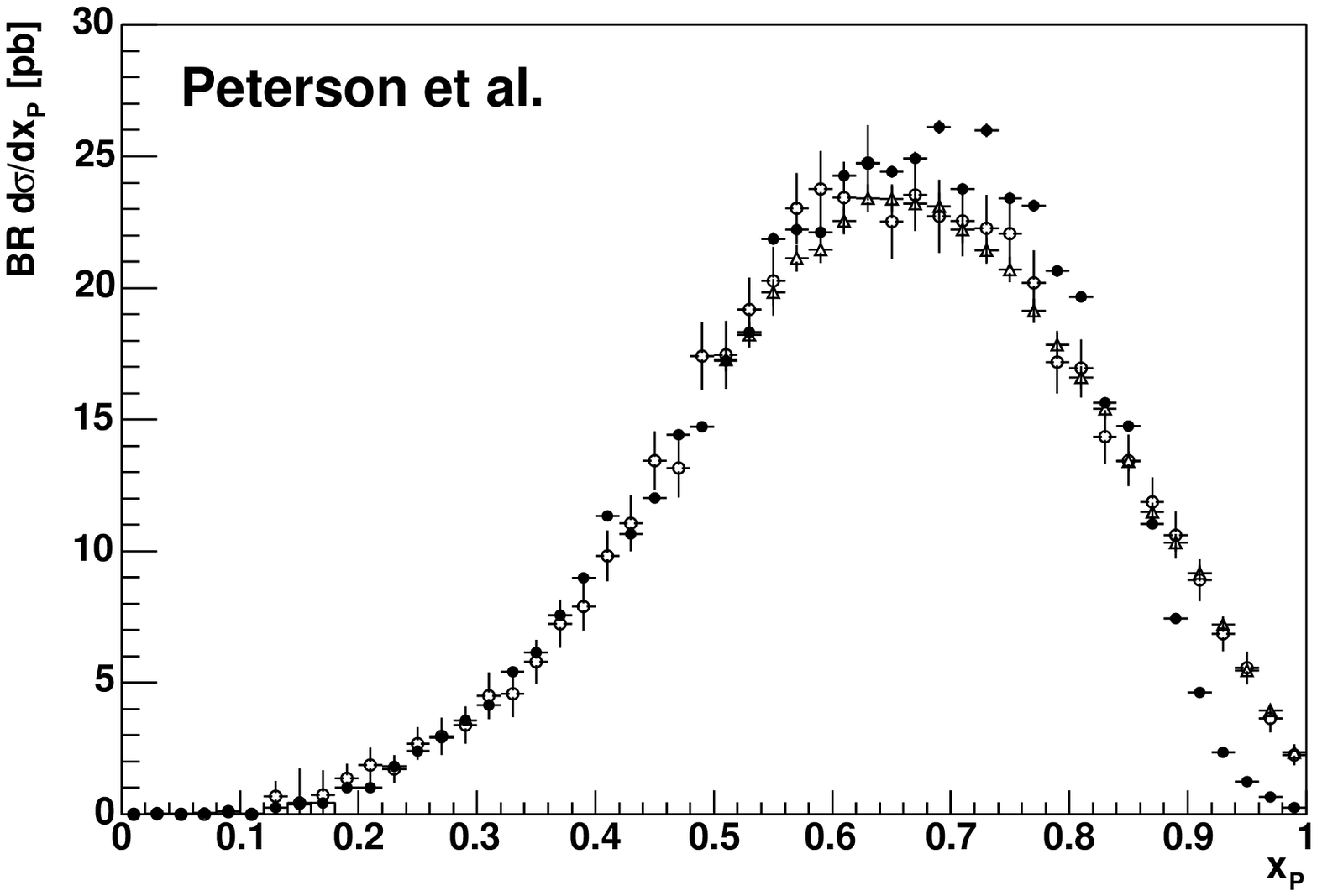}
}
\caption{
Efficiency corrected cross section BR$\times$d$\sigma$/d$x_p$ 
as a function of $x_p$ in comparison to fragmentation
models for $D^{*+}$ fragmentation 
(black circles: MC, 
white circles: on-resonance data, 
white triangles: off-resonance data). 
Only statistical errors are shown.
The parameters $a$,$b$ of the models
are optimized by a fit procedure in order to find the best agreement between data
and model. The deviations in the $x$$\simeq$1 region are an indication
for non-perturbative contributions.}
\end{figure}

\begin{table}[hhh]
\vspace*{0.5cm}
\tbl{The parameters of the fragmentation functions at
  the minimum of the $\chi^2/d.o.f.$ distributions,
which can be used as input to PYTHIA.}
{\begin{tabular}{@{}ccccccc@{}} \toprule
     & & $D^0$ & 
     $D^+$         & 
     $D^+_s$       & 
     $\Lambda^+_c$ & 
     $D^{*+}$      \\ \hline
%     & parameter
%     & par$s$ at min.
%     & par$s$ at min.
%     & par$s$ at min.
%     & par$s$ at min.
%     & par$s$ at min. \\ \hline
     Bowler & $a|b$ &
      0.12 $|$ 0.74 &
      0.12 $|$ 0.58 &
      0.12 $|$ 0.68 &
      0.34 $|$ 0.74 &
      0.22 $|$ 0.56 \\
      Lund & $a$ &
      0.26 &
      0.45 &
      0.2 &
      0.55 &
      0.58 \\
      Collins and Spiller & $\varepsilon_{c}'$ &
     0.04 &
     0.055 &
     0.04 &
     0.04 &
     0.075 \\
     Kartvelishvili & $\alpha_{c}$ &
     4.6 &
     4 &
     5.6 &
     3.6 &
     5.6 \\
     Peterson & $\varepsilon_{c}$ & 
     0.028 &
     0.039 &
     0.008 &
     0.011 &
     0.054 \\
     \botrule
   \end{tabular} \label{t_3}}
\end{table}

%%%

%Eq.~(\ref{diseqn})

%\begin{figure}[pb]
%\centerline{\psfig{file=ijmpaf1.eps,width=4.7cm}}
%\vspace*{8pt}
%\caption{This figure shows ... \label{f_..}
%}
%\end{figure}

%\begin{table}[ph]
%\tbl{This table shows ...
%}
%{\begin{tabular}{@{}cccc@{}} \toprule
%\botrule
%\end{tabular} \label{ta1}}
%\end{table}

%%journal paper
%\bibitem{jpap} R. Loren and D. B. Benson, {\it J. Comput. 
%System Sci.} {\bf 27}, 400 (1983).

%%collaboration
%\bibitem{colla} OPAL Collab. (G. Abbiendi {\it et al}.), 
%{\it Eur. J. Phys. C\/} {\bf 11}, 217 (1999).

%%to be published
%\bibitem{publ} R. Loren, J. Li and D. B. Benson, Deterministic
%flow-chart interpretations, to appear in {\it J. Comput. System Sci.} 

\end{document}